\documentclass[11pt,twoside]{article}
\usepackage{epsfig}
\usepackage{here}
\usepackage{amsmath}
      
\begin{document}

\begin{center}{\bf\Large  
Study of the ground and excited states of $\Lambda$ and
$\Sigma$ hyperon production at COSY
}
\end{center}
\vspace{0.5cm}
\begin{center}
   D. Grzonka \\[0.1cm]
   {\small \em 
         Institut f\"ur Kernphysik, Forschungszentrum J\"ulich, Germany
   }
\end{center}
\vspace{0.5cm}
\begin{center}
 \parbox{0.9\textwidth}{
  \small{
    {\bf Abstract:}\
COSY with a maximum beam momentum of $\sim$~3.65 GeV/c allows 
the production of ground state ($\Lambda$ and $\Sigma$) and
excited hyperons ($\Lambda$(1405), $\Sigma$(1385) and
$\Lambda$(1520) ) in elementary $NN$ interactions.
The existing data base in this field is rather poor.
A systematic study of the hyperon production at COSY
will result in an improvement of our understanding concerning
topics like hyperon--nucleon interaction, kaon--nucleon interaction,
nucleon resonances, strangeness production mechanism and structure of
the $\Lambda$(1405).
Precise data on $\Lambda$ and $\Sigma$ production have been 
produced at the COSY-11 and TOF installation.
With the WASA detector at COSY 
these studies can be continued and extended to channels 
including photons in the final state.
     }
 }
\end{center}

\vspace{0.5cm}
\section{Introduction}

The cooler synchrotron COSY with a maximum beam momentum of $\sim$~3.65 GeV/c
allows the production of $\Lambda$ and $\Sigma$ hyperons 
in elementary NN-interactions via the associated strangeness production
$NN \rightarrow N Y K$.
Ground state ($\Lambda$, $\Sigma^-$, $\Sigma^0$, $\Sigma^+$)
as well as the excited hyperon states $\Lambda$(1405), $\Sigma$(1385) and
$\Lambda$(1520) are accessible.

The physics motivation for the production of ground state hyperons
is very different from the motivation to study their excited states.
The ground state hyperons are well defined (qqq) states of the $J^P \, = \, 1/2^+$
baryon octet with S=-1 and its properties like mass and decay modes are 
very well known.
Production studies of the ground state hyperon aims at 
information on the 
production mechanism and on the interaction with hadrons.
Concerning the excited hyperons especially the $\Lambda$(1405) 
the main focus is on the resonance shape and its structure.
It is under discussion if or how much of the $\Lambda$(1405) 
can be attributed to a 3-quark state or a 
meson-baryon molecule.

From the experimental point of view the studies of ground and excited hyperons
are comparable. In both cases delayed decays
are involved which allows very efficient trigger generation
and event identification. The multiplicity and particle
configuration to be detected in the final
states is the same.

In table \ref{hyperon} and \ref{hyp*} some properties of the hyperon states 
relevant for the experimental studies at COSY are summarized \cite{pdg04}.

\begin{table}
\begin{center}
\begin{tabular}{lrrrrr}
\hline
state & $I(J^P)$& M  [MeV]& $c\tau$ & decay modes\\
\hline
\hline
$\Lambda$ & $0(\frac{1}{2}^+)$ & 1115.683$\pm$0.006 & 7.89 cm&
$p\pi^-$(64\%), $n\pi^0$(36\%) \\
\hline
$\Sigma^+$ &1($\frac{1}{2}^+$)&1189.37  $\pm$ 0.07   &2.404 cm&$p
\pi^0$(52\%), $n \pi^+$(48\%)\\
\hline
$\Sigma^0$ & 1($\frac{1}{2}^+$) & 1192.642$\pm$0.024 & 2.22  \, 10$^{-11}$ cm & $\Lambda \gamma$($\sim$100\%) \\
\hline
$\Sigma^-$ & 1($\frac{1}{2}^+$) & 1197.449$\pm$0.03 & 4.434 cm& $n \pi^-$ ($\sim$100\%) \\
\hline
\end{tabular}
\caption{Properties of $\Lambda$ and $\Sigma$ ground state hyperons 
\cite{pdg04}.}
\label{hyperon}
\end{center}
\end{table}

\begin{table}
\begin{center}
\begin{tabular}{lrrrrr}
\hline
state & $I(J^P)$& M  [MeV]& $\Gamma$ [MeV]& decay \\
\hline
$\Lambda(1405)$  & 0($\frac{1}{2}^-$)&1406$\pm$ 4 & 50.0$\pm$2.0& $\Sigma \pi$(100\%) \\
\hline
$\Lambda(1520)$&0($\frac{3}{2}^-$)&1519.5$\pm$1.0& 15.6$\pm$1.0&
$N\bar{K}$(45\%),$\Sigma \pi$(42\%) \\
               &                  &      & & $\Lambda \pi \pi$(10\%) \\    
\hline
$\Sigma^+$(1385) & 1($\frac{3}{2}^+$) & 1382.8$\pm$0.4 & 35.8$\pm$0.8  &\\
$\Sigma^0$(1385) & 1($\frac{3}{2}^+$) & 1383.7$\pm$1.0 & 36$\pm$5&$\Lambda \pi$(88\%),$\Sigma \pi$(12\%) \\
$\Sigma^-$(1385) & 1($\frac{3}{2}^+$) & 1387.2$\pm$0.5 &39.4$\pm$ 2.1 & \\
\hline
\end{tabular}

\caption{Properties of excited hyperon states relevant for
  experimental studies at COSY \cite{pdg04}
.}
\label{hyp*}
\end{center}
\end{table}

\section{Ground state hyperons}
A systematic study of the ground state hyperon production will
result in an extended data base for the hyperon-nucleon interaction
which is an important ingredient in different
physics questions. The dynamics of few body systems containing
strangeness is not well known. Models describing hyperon--nucleon
scattering \cite{maessen,juel,juel1,rijken,jueln}
rely on flavour SU(3) symmetry to fix the baryon--meson
couplings and the remaining parameters are fitted to the data. 
The existing data on hyperon--nucleon scattering
are insufficient to prove the validity of this procedure. 
A better understanding of the hyperon--nucleon interaction 
has an impact on various topics
like the formation of hyper-nuclei
\cite{nogga} or the structure of neutron stars \cite{neutronstars}. 

\begin{figure}[H]
\begin{center}
\epsfig{file=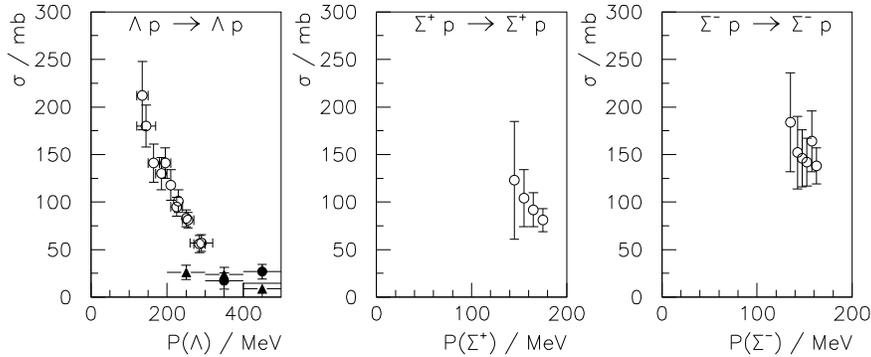,width=\textwidth}
\hfill
  \caption{\label{grzonkafig1} \small 
Available data on low energy elastic hyperon scattering $Y p \rightarrow Y p$
for $\Lambda$ (open circle \cite{ale68},~filled circle \cite{hau77},
triangle \cite{kad71})
, $\Sigma ^+$  (\cite{eis71}) and 
$\Sigma ^-$ (\cite{eis71}) hyperons.
}
\end{center}
\end{figure}

Data on $\Lambda$ and $\Sigma$ ground state hyperon production 
related to hyperon--nucleon interaction studies mainly result from
hyperon-- nucleon scattering extracted from bubble chamber experiments.
The hyperon production was induced by a $K^-$ beam via 
$K^- + p \rightarrow Y \pi$ where Y is a $\Lambda$ or a $\Sigma$
hyperon. These $\Lambda$ and $\Sigma$
hyperons then may scatter elastically on another proton of the hydrogen bubble
chamber.
Experiments have been performed at CERN \cite{ale68, eis71, sec68,
eng66, kad71} or at SLAC \cite{hau77}.
Scattering data for higher $\Lambda$ momentum (1-17 GeV/c)  have been
produced at BNL by using a proton beam \cite{and75}.
The available data on low energy elastic hyperon scattering are shown
in figure \ref{grzonkafig1}. 

Most of the data are available for the $\Lambda$-p system 
but also here the experimentally extracted  
low energy parameters of s-wave scattering,  
the scattering length and the effective range, have large error bars.
A detailed analysis of the 
world data set for elastic $\Lambda p$ scattering result in 
$a_s=-1.8+(+2.3-4.2) \ \mbox{fm and }  
a_t=-1.6+(+1.1-0.8) \ \mbox{fm}$ \cite{ale68} 
for the spin singlet and spin triplet scattering lengths, respectively, 
where the errors are strongly correlated.
Furthermore a clear separation of the spin singlet and triplet channel    
is not possible from the data.
An analysis within an effective range approximation has been performed
also for the $\Sigma^+$ elastic scattering data \cite{eis71} but
due to the poor data base 
only a  wide region for the scattering lengths could be extracted
by applying some model based relations
for the scattering lengths with
$-6fm < a_s < +2.5 fm$ and  $-2 fm < a_t < 1.3 fm$ 
in which singlet and triplet scattering lengths
are correlated.
These secondary scattering experiments are limited at the low momentum
end due to the hyperon decay and the necessity to have a detectable track 
of the scattered proton.
Therefore the data have to be extrapolated to the threshold resulting
in a large error width.

The data on elastic hyperon scattering will be extended by
measurements at KEK where first results on differential observables
are published \cite{ahn99, kon00}. 
The hyperons are produced via the ($\pi^+ / \pi ^- , K^+$) reactions and the
detection
system consists of
a scintillating fiber arrangement 
\cite{ahn99, iei01, kon00} or a 
``Scintillating track image camera'' \cite{asa03}
where the tracks of charged
particles in a scintillator are viewed by ccd cameras.

An alternative way to study the nucleon--hyperon interaction are
production reactions with a nucleon and a hyperon in a multi 
particle exit channel. First data in this field were extracted
from the reaction $K^-d\to \pi^-p\Lambda$  \cite{tan69}, 
leading to a scattering length of $-2\pm 0.5$ fm 
via fitting the invariant mass distribution to an effective range 
expansion. The author argued that this value is to be interpreted as the spin 
triplet scattering length but it is difficult to estimate the theoretical 
uncertainty, the error given includes the experimental uncertainty
only.

Also at COSY a first attempt to determine the low energy parameters of 
the $\Lambda N$ interaction via the reaction 
$pp\to K^+p\Lambda$ has been performed \cite{bal98b}.
We extracted an average value of $-2\pm 
0.2$ fm for the $\Lambda N$ scattering length in a Dalitz plot
analysis that utilizes the 
effective range expansion. The effective range parameters $a$ and $r$
are strongly correlated and for their determination the $\Lambda N$ elastic cross sections
data had to be included in the analysis to determine the parameters.
Furthermore the spin singlet and triplet Y-N states
contributes to the data i.e. only a spin averaged value was extracted.

In Ref. \cite{gib73} it was suggested to use the reaction $K^-d\to 
n\Lambda \gamma$, where the initial state is in an atomic bound state, to 
determine the $\Lambda N$ scattering lengths. From the experimental side so 
far, a feasibility study was performed which demonstrated that a separation of 
background and signal is possible \cite{kdexp}.  The reaction $K^-d\to 
n\Lambda \gamma$ was studied theoretically in more detail in 
\cite{akh78,wor90,gib00}. The main results especially of the last work are that 
it is indeed possible to use the radiative $K^-$ capture to extract the 
$\Lambda N$ scattering lengths and that polarization observables could be used 
to disentangle the different spin states.

\begin{figure}
\begin{center}
\epsfig{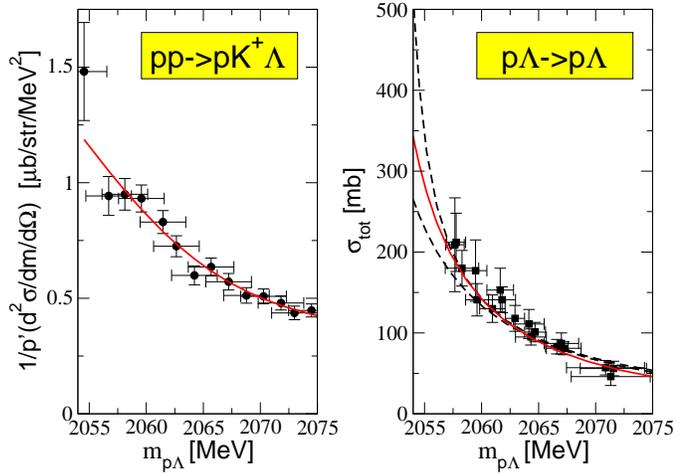}
\end{center}
\caption{\small{Left panel: $pp \to K^+X$ inclusive missing mass spectrum at
$T_p$ = 2.3 Gev corrected by the two body
phase space. Right panel: world data set on total cross sections for $\Lambda
N$ elastic
scattering. In both panels the solid curve represents a best fit to the data. In 
the right panel two more fits still consistent with the data were added
to show the uncertainty in the extrapolation.}}
\label{grzonkafig2}
\end{figure}

In a 3 particle exit channel like $pp \rightarrow pK^+ \Lambda$ 
the relative momentum in each individual 2 particle subsystem 
ranges from the maximum momentum given by the kinematics down to 0.
Therefore event samples can be selected with
very low relative hyperon--nucleon momentum down to the threshold 
which is not accessible by conventional elastic scattering.
In figure \ref{grzonkafig2} \cite{han04}
the situation is illustrated in a comparison of the $\Lambda$p
scattering data with a data set of inclusive $K^+$ production
via $pp \rightarrow K^+ X$
measured at SATURNE \cite{sie94}. Below the
$\Sigma ^0$ production threshold the dominant 
reaction channel is $pp \rightarrow p K^+ \Lambda$. 
Neglecting other reaction channels
like $pp \rightarrow p K^+ \Lambda \gamma$ the kinetic energy 
of the $K^+$ is directly related to $m_{\Lambda p}$.
In figure \ref{grzonkafig2} it is clearly shown that the $m_{\Lambda p}$
distribution is covered by data points up to the threshold. 
The extraction of the scattering length by fitting the data with
theoretical curves is much more precise than the extrapolation of the
scattering data. The different curves fitted to the scattering data 
cover the present error range for the scattering length.
The highest sensitivity is at smallest   $m_{\Lambda p}$ values where
no data are available.
These inclusive $K^+$ data are not really useful to improve the
knowledge of the $\Lambda p$ scattering length because singlet and
triplet state are not separated. This example should only demonstrate
the advantages of the $pp \rightarrow pK^+ \Lambda$ reaction for such
studies. 
Another important feature of the $pp \rightarrow pK^+ \Lambda$
reaction is the high momentum transfer in the reaction.
High momentum transfer is correlated with
a production process of short-range nature \cite{gas04} 
which is insensitive to details of the production mechanism
resulting in reliable error estimates in the theoretical treatment.

The curves in figure \ref{grzonkafig2} result from a new method 
to determine the $\Lambda p$ scattering lengths proposed in
Ref. \cite{gas04}. It allows the extraction of the $YN$ scattering lengths from the
production data directly.  In particular, an integral representation for the
$\Lambda N$ scattering lengths in terms of a differential cross section of
reactions with large momentum transfer such as $pp\to K^+p\Lambda$ or $\gamma
d\to K^+n\Lambda$ was derived.  This formula should enable the determination
of the scattering lengths to a theoretical uncertainty of at most 0.3
fm.
Experimentally production data of sufficient accuracy can be achieved
as is seen by the SATURNE data  \cite{sie94} introducing an
experimental uncertainty in the extraction of the scattering length of
0.2 fm.

In Ref. \cite{gas04} it was also shown that already a measurement of a single
spin asymmetry in $\vec pp\to YNK$ enables to isolate the spin triplet 
contribution from the final $YN$ state, since the Pauli Principle strongly
limits the number of structures allowed for the initial state. It was
especially shown that
$$
\frac{d^2\sigma (\uparrow)}{dm' \, ^2dt}-\frac{d^2\sigma (\downarrow)}{dm' \, ^2dt}
$$
gets contributions from the spin triplet final state only, as long as the kaon
in the final state is emitted at $90^o$ in the cm system and the outgoing
$YN$ is in an $S$--wave. 
The arrows in the above expression
denote the polarization of either the beam or the target that is to be chosen
perpendicular to the beam axis.

The method in Ref. \cite{gas04} was derived under the assumption that there is no inelastic
channel available for the hyperon--nucleon system. This condition only holds
for the $\Lambda N$, $\Sigma^+p$ as well as the $\Sigma^-n$ channel (where for
an analysis of the second channel the Coulomb interaction needs to be included 
in the formalism---work on this straight forward extension of the formalism is in
progress). These $\Sigma N$ channels are purely isospin $3/2$ states and thus
do not couple to the $\Lambda N$ channel. The other $\Sigma N$ channels,
however, are inelastic. As a consequence their scattering lengths aquire
imaginary parts and the dispersion analysis in Ref. \cite{gas04}
can no longer be applied.

The future studies of the hyperon production at COSY 
induced by NN interactions 
via $NN \rightarrow N K Y$ will
significantly extend and improve the data base on the
hyperon--nucleon interaction.

But apart from the hyperon--nucleon interaction there is much more
physics which can be extracted from these data.
First of all the production mechanism can be studied. 
Dalitz plot analysis of $pp \rightarrow pK^+ \Lambda$ data
taken at COSY by the TOF collaboration showed that the lowest nucleon 
resonances which couple to the $KY$ channel are a dominant
strongly energy dependent 
contribution to the hyperon production \cite{eyr03}.
At a beam momentum of 2.85 GeV/c \mbox{(Q = 171 MeV)} only the N*(1650) is
relevant, at 2.95 GeV/c \mbox{(Q = 204 MeV)} also the N*(1710) gives a significant 
contribution and at 3.2 GeV/c \mbox{(Q = 284 MeV)} both, N*(1650) and N*(1710), contribute
with comparable strength.
Close to threshold data have been supplied by the COSY-11
collaboration
\cite{bal98, sew99, kow04}
resulting in information on the hyperon-nucleon interaction \cite{bal98b}
as discussed above but especially the comparison
of threshold production of different hyperon channels like
the $\Lambda /\Sigma ^0$ comparison \cite{kow04, roz04}.
will help to disentangle the production mechanisms close to threshold. 
Spin transfer coefficients by using a polarized proton beam and
measuring the $\Lambda$ polarization 
$\vec{p}p \rightarrow p K^+ \vec{\Lambda}$
have been produced by the DISTO collaboration at a beam momentum of
3.67 GeV/c  \mbox{(Q = 430 MeV)} \cite{bal99}.
The model dependent interpretation of the data indicates a production
mechanism dominated by kaon exchange.
 
With the hyperon production data available up to now first indications 
of the relevant production mechanisms are available but
much more informations are still needed.

The kinematically complete measurement of the
3 particle $NKY$ exit channel allows a
detailed investigation of all individual
2-body subsystems.
Besides the $YN$ system also the
$NK$ as well as the $KY$ system can
be studied.
The $KY$ system is mainly important for the investigation of 
nucleon resonances, especially the $N^* (1650)$, $N^* (1710)$ and 
$N^* (1720)$ which have large branching ratios into the $\Lambda K$
channel.
In the $NK$ subsystem the 
interaction between nucleons and K mesons
can be observed. Of special interest here is the 
$\Theta ^+$ resonance which was observed by many experiments and is
interpreted as a penta-quark system \cite{pdg04b} 
proposed by Diakonov et al. \cite{dia97}.
A clear signal of the $\Theta ^+$ has also been detected 
in the $K^0 p$ system
by the COSY TOF collaboration \cite{eyr04}. 
Further measurements at TOF with improved statistics are on the way 
and a double polarization experiment to determine the parity of
the $\Theta ^+$ \cite{han04b} is in preparation.

\section{Excited hyperons}
The most interesting topic in the excited hyperon sector is the 
shape and the structure of
the $\Lambda (1405)$ resonance. Its a long standing discussion if the $\Lambda
(1405)$ is a 3-quark state or a $\bar{K}N$ / $\pi \Sigma$ bound state,
see \cite{kim00} and references cited therein.

In a recent study of the poles in the meson-baryon scattering matrix
in a chiral unitary approach some resonances are 
dynamically generated close to the $\Lambda (1405)$ mass \cite{jid03}.
It is argued that the $\Lambda (1405)$ is not only one but a
superposition of two close
resonances which could be experimentally separated due to the
different couplings to the $\pi \Sigma$ and $\bar{K} N$ states.
Whereas the resonance at lower mass couples stronger to the $\pi \Sigma$ state
the resonance at higher mass couples mostly to the 
$\bar{K} N$ state.
Similar results have been obtained in other studies, see \cite{kim00, jid03} and the references cited therein.
But a definite conclusion about the structure of the  $\Lambda (1405)$
can not be drawn from the existing data.

The data in the excited hyperon sector mostly result from production studies in bubble
chambers via $\pi^- p$ or $K^- p$ interactions \cite{hem84, tho73,
bau84, agu81, bar79, pdg04b}.
Recently the photoproduction of the $\Lambda (1405)$ has been studied
at SPring-8/LEPS via $p(\gamma , K^+ ) Y^*$ \cite{ahn03}. In the missing mass
distribution the $\Lambda (1405)$ is overlapping with the $\Sigma (1385)$.
A separation is achieved in the ($\pi \Sigma$) invariant mass
distribution because the $\Lambda (1405)$ decays with  100\% into the
$\pi \Sigma$ channel but the  $\Sigma (1385)$ has a branching of only
12\% into this channel.
The comparison of the $\pi^+ \Sigma ^-$ and the $\pi^- \Sigma ^+$ 
invariant mass distributions shows
different line shapes which could be an indication that the $\Lambda (1405)$ 
is a meson-baryon bound state.

At COSY the production of overlapping $\Lambda ^* (1405)$ and 
$\Sigma ^* (1385)$ hyperons have been observed in
$pp$ interactions \cite{que01, bue03}. But for a detailed study
first of all a separation of these resonances by measuring their decay
products is necessary.

\section{Experimental study of hyperons at COSY}

The experimental details to study the ground and excited hyperons are
very similar. Up to now the hyperon production at COSY has been
investigated at the COSY-11 and at the TOF installation.

COSY-11 \cite{bra96} is an internal experiment operating with a beam of hydrogen
clusters as target in front of a COSY machine dipole used as a
magnetic spectrometer for the charged reaction products. The tracks
and velocity of positively charged reaction products are measured
resulting in the determination of its 4-momentum vectors. This system is limited
in acceptance but achieves a high momentum resolution of the measured
particles. Hyperon production is studied via $pp \rightarrow N K^+ Y$
where the hyperons are identified via
the missing mass method.
Due to the high missing mass resolution the hyperons can be clearly
identified on a rather low background level.
Excitation functions from about 1 - 60 MeV excess energy have been measured for
$pp \rightarrow p K^+ \Lambda$ and $pp \rightarrow p K^+ \Sigma ^0$ 
and data of the $pp \rightarrow n K^+ \Sigma ^+$ channel are under
analysis.
For the $pp \rightarrow p K^+ \Lambda$ reaction also a Dalitz plot analysis has
been performed resulting in the determination of a spin averaged 
$\Lambda p$ scattering length.
These studies will be continued by a measurement with polarized beam.
In the analysis the method of Gasparian et al. \cite{gas04} will be used
to isolate the spin triplet scattering length which requires the
measurement of kaons emitted in $90^o$ in the cm system. 
Although COSY-11 is a magnetic spectrometer device the special
configuration of a C-type magnet, open to the detector side with a thin
exit window for the reaction products, allows to measure the full
scattering angle distribution for a thin slice in the beam plane.
Therefore kaons emitted around $90^o$ cms can be detected.
The proposal for this experiment has been accepted and the measurement
is planned for spring 2005 \cite{grz04}.

The TOF spectrometer \cite{tof04} is a large acceptance external experiment.
A small liquid hydrogen cell (a few tens $mm^3$ volume) is used as a
target surrounded by a first layer of scintillators followed by a
second layer in a distance of a few m dependent on the actual setup.
A special decay spectrometer is installed close to the target with
several layers of Si-$\mu$strip and scintillating fiber detectors.
This decay spectrometer allows to trigger on delayed decays and
measures the tracks of charged decay products.
The event identification is mainly done by geometry which is
sufficient if a precise definition of target vertex and hit
positions is realized as it is the case at TOF.
Data have been produced for $pp \rightarrow p K^+ \Lambda$ 
for Q values between 55 MeV and 284 MeV resulting in Dalitz plots 
with a full coverage of the phase space \cite{eyr03}. Due to the decay asymmetry of
the weak $\Lambda$ decay also the $\Lambda$ polarization could be
determined by measuring the tracks of the charged decay particles.
Furthermore the reactions  $pp \rightarrow p K^+ \Sigma ^0$, 
$pp \rightarrow p K^0 \Sigma ^+$ and $pp \rightarrow n K^+ \Sigma ^+$
have been studied.

In table \ref{hypreac} all ground state hyperon production channels
accessible in elementary NN interactions at COSY are listed.
 
\begin{table}
\begin{center}
\begin{tabular}{lll}
\hline
primary reaction & final state & br. \\
\hline
$ pp \rightarrow p K^+ \Lambda  $ & $p K^+ p \pi^-  $ & 0.64\\
& $p K^+ n \pi^0  $ $\rightarrow p K^+ n \gamma \gamma $& 0.36\\
\hline
$pp \rightarrow p  K^+ \Sigma ^0$ & 
$p K^+ \Lambda \gamma $ $\rightarrow p  K^+ p \pi^- \gamma$& 0.64\\
&  $p   K^+ \Lambda \gamma$ $\rightarrow p  K^+ n \pi^0 \gamma$
$\rightarrow p  K^+ n \gamma \gamma \gamma$ & 0.36\\
\hline
$pp \rightarrow p K^0 \Sigma ^+ $ &
$p \pi^0 \pi^0 p \pi^0 $
$ \rightarrow p \gamma\gamma \, \gamma\gamma p \gamma\gamma \, $& 0.08\\
  & $p \pi^+ \pi^- p \pi^0  $
$ \rightarrow p  \pi^+ \pi^- p \gamma\gamma \,  $& 0.18\\

  & $p  \pi^0 \pi^0 n \pi^+ $ 
$ \rightarrow p\gamma\gamma \, \gamma\gamma  n \pi^+  $ & 0.08\\
 & $p  \pi^+ \pi^- n \pi^+ $  & 0.16\\
\hline
$ pp \rightarrow n K^+ \Sigma ^+ $ &
$ n K^+   p \pi^0 $ $\rightarrow n  K^+ p  \gamma\gamma$ & 0.52\\
 &
$n K^+  n \pi^+  $ & 0.48\\
\hline
$ pn \rightarrow n K^+ \Lambda  $ & $n K^+ p \pi^-  $ & 0.64\\
 & $n K^+ n \pi^0  $
$\rightarrow n  K^+  n \gamma\gamma  $ & 0.36\\
\hline
$ pn \rightarrow n K^+ \Sigma^0  $ & $n K^+  \Lambda \gamma$
$\rightarrow n K^+ p \pi^- \gamma  $ & 0.64\\
 & $n K^+  \Lambda \gamma$   $\rightarrow n K^+ n \pi^0 \gamma  
$ $\rightarrow n K^+ n \gamma \gamma \gamma  $& 0.36\\
\hline
$pn \rightarrow n K^0 \Sigma ^+ $ &
$n \pi^0 \pi^0 p \pi^0 $
$ \rightarrow n \gamma\gamma \, \gamma\gamma p \gamma\gamma \, $& 0.08\\
  & $n \pi^+ \pi^- p \pi^0  $
$ \rightarrow n  \pi^+ \pi^- p \gamma\gamma \,  $& 0.18\\
  & $n  \pi^0 \pi^0 n \pi^+ $ 
$ \rightarrow n\gamma\gamma \, \gamma\gamma  n \pi^+  $ & 0.08\\
 & $n  \pi^+ \pi^- n \pi^+ $  & 0.16\\
\hline
$ pn \rightarrow p K^0 \Lambda  $ & $p \pi^0 \pi^0 p \pi^-  $
$\rightarrow p \gamma\gamma \, \gamma\gamma p \pi^-  $ & 0.1\\
& $p \pi^+ \pi^- p \pi^-  $ & 0.22\\
   & $p \pi^0 \pi^0 n \pi^0  $
$\rightarrow p \gamma\gamma \, \gamma\gamma n  \gamma\gamma  $ & 0.06\\
& $p \pi^+ \pi^- n \pi^0$ $\rightarrow p \pi^+ \pi^-  n \gamma\gamma    $ & 0.12\\
\hline
$ pn \rightarrow  p K^0 \Sigma^0  $ & $p \pi^0 \pi^0 \Lambda \gamma  $
$\rightarrow p \gamma\gamma \, \gamma\gamma p \pi^- \gamma$ & 0.1\\
 & $p \pi^+ \pi^-  \Lambda \gamma  $   $\rightarrow p \pi^+ \pi^- p \pi^- \gamma $ & 0.22\\
 & $p \pi^0 \pi^0 \Lambda \gamma  $
$\rightarrow p \gamma\gamma \, \gamma\gamma n \gamma\gamma \gamma$ & 0.06\\
 & $p \pi^+ \pi^-  \Lambda \gamma  $  $\rightarrow p \pi^+ \pi^- n \pi^0 \gamma $
$\rightarrow p \pi^+ \pi^- n \gamma\gamma \,\gamma $ & 0.12\\
\hline
$ pn \rightarrow p K^+ \Sigma^-  $ & $p K^+ n \pi ^-$& 1.0\\
\hline
\end{tabular}
\label{hypreac}
\end{center}
\end{table}

For a systematic detailed study of the hyperon production the full
information of the measured events including the 4-momentum vectors 
of the decay particles
is needed. This will allow to extract nearly background free event
samples required for Dalitz plot analysis.
Some reaction channels in table \ref{hypreac} could be studied at TOF
as is shown by the produced data but at TOF no photon detector is installed.
In most channels one or more $\gamma$'s are included in the exit
channel. Therefore the WASA detector \cite{cal02, zab02} 
which will move to COSY in the near future seems to be
well suited for these hyperon production studies.
WASA includes a detection system for charged particles as well as an
electromagnetic calorimeter for the detection of photons.

\begin{figure}
\begin{center}
\epsfig{file=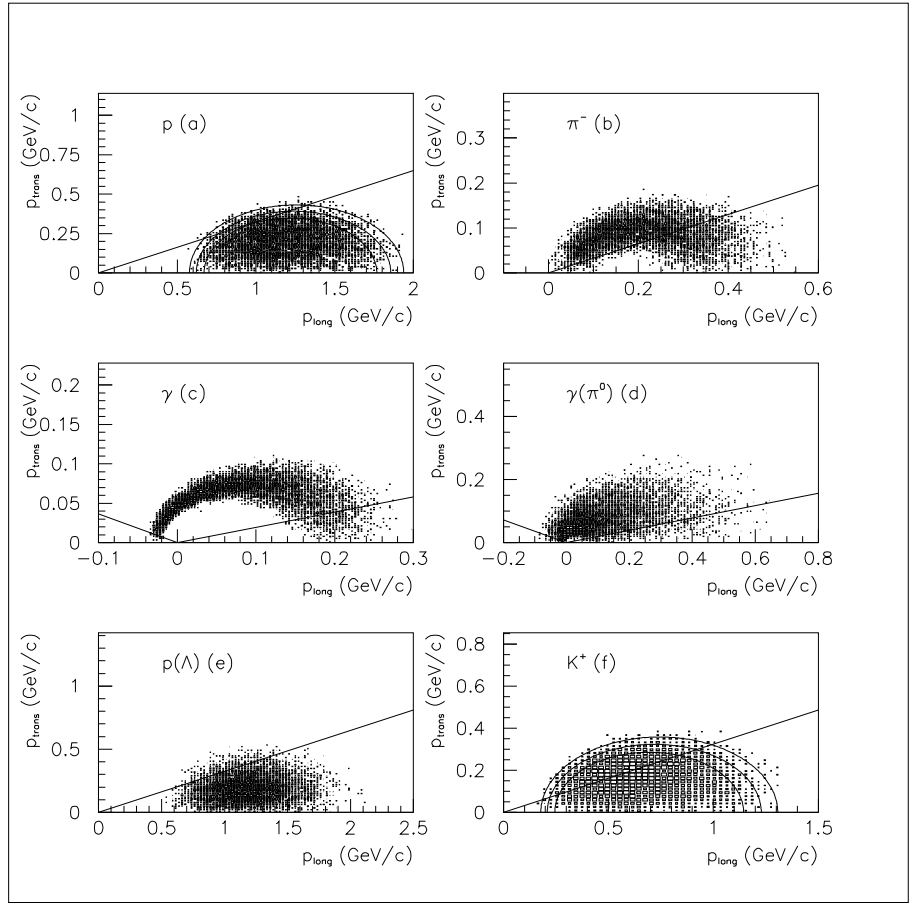, width=0.9\textwidth}
\end{center}
\vspace*{-0.5cm}
\caption{\small{
Momentum distribution of the final state particles for the reaction
channel :$pp \rightarrow \Lambda (1405) p K^+ \rightarrow 
p^{(a)} \pi^{-(b)} \gamma^{(c)} \gamma
\gamma^{(d)} p^{(e)} K^{+(f)}$.
The ellipses show the kinematic limits for a $\Lambda (1405)$ production
with PDG mass $\pm$ 25 MeV.
The straight lines indicate the acceptance region covered by the em calorimeter of the WASA detector.
}}
\label{grzonkafig3}
\end{figure}

With the possible production channels of the excited hyperons:\\
$pp \rightarrow \Lambda ^* p K ^+ \ / \ \Sigma ^{*+} p K ^0 \ / \ 
\Sigma ^{*0} p K ^+$ and
$pn \rightarrow \Lambda ^* p K ^0 \ / \ \Sigma ^{*0} p K ^0 \ / \ 
\Sigma ^{*-} p K ^+$
\\
a corresponding table for the exit particle configuration can be prepared
with the dominant $Y^*$ decays given in table \ref{hyp*}.
The final states are similar to the ground state hyperons with a
particle multiplicity of about 6 including charged particles and
photons.
As an example
in fig. \ref{grzonkafig3} the momentum distribution of the final state
particles in the reaction channel
$pp \rightarrow \Lambda (1405) p K^+ \rightarrow p \pi^- \gamma \gamma
\gamma p K^+$ is shown for a Monte Carlo generated event sample.
The straight lines in the plots indicate the acceptance of the WASA calorimeter
for photon detection. The charged particles are mainly emitted in
forward direction which requires a high resolution tracking system in
this area.  
How well the existing WASA detection system will be suitable for these
hyperon production studies and which modifications would be necessary
has to be studied via Monte Carlo event samples.
A close to target tracking system, not existing in the present WASA
setup,
to identify the delayed decay
vertices is for sure indispensable to select clean event samples.
This point is discussed in letters of intend for possible experiments
with WASA at COSY \cite{loi04}.

\section{Summary}
A systematic study of hyperon production at COSY in the 3-body final state 
$NN \rightarrow N K Y ^{*}$
will result in detailed information on various topics.
By measuring all final state particles including the decay products
nearly background free event samples of the various reaction 
channels can be selected which is important for efficient Dalitz plot
analysis. With the use of a polarized beam (and target) and the
measurement of the hyperon polarization (at least for $\Lambda$ and
$\Sigma ^+$ with their high decay asymmetry) very selective studies
will be possible. 

By selecting appropriate kinematic regions 
-- difficult or even impossible to access 
by conventional scattering experiments -- the interactions of the 
individual subsystems
($NY$), ($NK$) and ($KY$) can be analyzed.
Precise data will result on the hyperon--nucleon interaction, requested for a better
understanding of the dynamics in baryon systems with strangeness,
the kaon--nucleon system, of special interest in view of the recent observations of
the $\Theta ^+$ resonance and the kaon--hyperon system, important for
the investigation of nucleon resonances. 
In case of the excited hyperons the knowledge about the structure of
these resonances will be much improved.
Furthermore with these detailed studies the mechanisms of hyperon
production can be clarified.

By their results on hyperon production the running experiments
have proven that high precision data in this field are possible
at COSY. Valuable results have been obtained but much more information is
necessary in order to fully understand the strangeness sector.

\section*{Acknowledgments}

 The work has been supported by the European Community - Access to
Research Infrastructure action of the
Improving Human Potential Programme
and by the DAAD Exchange Programme (PPP-Polen).


\end{document}